\documentclass{jaa}
\usepackage{graphicx}
\usepackage{multirow}
\usepackage{booktabs}
\usepackage{xcolor}
\usepackage{hyperref}

\begin{document}\sloppy

\title{\textit{AstroSat} view of the NLS1 galaxy Mrk~335
}

\author{Savithri H. Ezhikode\textsuperscript{1}, Gulab C. Dewangan\textsuperscript{1} and Ranjeev Misra\textsuperscript{1}}
\affilOne{\textsuperscript{1}Inter-University Centre for Astronomy and Astrophysics, Post Bag 4, Ganeshkhind, Pune 411007, India\\}

\twocolumn[{

\maketitle

\corres{savithri@iucaa.in}

\msinfo{: 07 Nov 2020}{: 24 Dec 2020}

\begin{abstract}

We present the results from the multi-wavelength monitoring observations of the Narrow-Line Seyfert 1 galaxy Mrk~335 with \textit{AstroSat}. We analysed both the X-ray (SXT \& LAXPC) and UV (UVIT) data of the source at two epochs, separated by $\sim 18$~days. The source was in a low flux state during the observations, and the X-ray spectra were found to be harder than usual. The presence of soft X-ray excess was identified in the observations, and the broadband X-ray continuum was modelled with power-law and blackbody (modified by intrinsic absorption) and a distant neutral reflection component. We did not find any variability in the X-ray spectral shape or the flux over this period. However, the UV flux is found to be variable between the observations. The obtained results from the X-ray analysis point to a scenario where the primary emission is suppressed and the component due to distant reflection dominates the observed spectrum.
\end{abstract}


}]

\doinum{}
\artcitid{\#\#\#\#}
\volnum{000}
\year{0000}
\pgrange{1--}
\setcounter{page}{1}
\lp{1}

\vspace*{1cm}

\section{Introduction}
\label{intro}

Narrow-line Seyfert 1 galaxies (NLS1s) are a particular class of active galactic nuclei (AGN) with some extreme properties. They show strong Fe-II emission in the optical band along with narrow permitted lines and weak [OIII] emission (Osterbrock \& Pogge, 1985; Goodrich 1989). In the X-ray band, they are generally characterised by enhanced spectral and flux variability and the presence of soft excess emission (e.g. Boller et al. 1996).

Mrk~335 (RA=00h06m19.5s, DEC=+20d12m11s) is an NLS1 galaxy at redshift=0.026. The source is well known for showing dramatic fluctuations between high and low flux states in the X-ray band. Mrk~335 was observed with earlier observatories like \textit{Uhuru}, \textit{Einstein}, \textit{EXOSAT}, \textit{Ginga}, \textit{ROSAT}, \textit{ASCA}, and \textit{BeppoSAX} when it was an X-ray bright source in the sky (eg. Tananbaum et al. 1978; Halpern 1982). Its intensity dropped from the brightest stage to the very low flux state in 2007 (Grupe et al. 2007). Later, the source remained mostly in a low flux state though it has been reported to showing episodes of X-ray flaring activities (e.g. Gallo et al. 2018).

Mrk~335 was extensively studied in the optical/UV and X-rays (Grupe et al. 2007, 2008, 2012; Longinotti et al. 2013; Gallo et al. 2013; Parker et al. 2014; Komossa wt al. 2014; Chainakun \& Young 2015; Keek \& Ballantyne 2016, Sarma et al. 2015, Gallo et al. 2015, Wilkins et al. 2015). The X-ray spectra obtained with \textit{Ginga}, \textit{ASCA}, \textit{BeppoSAX}, \textit{XMM-Newton} and \textit{NuSTAR} showed evidences for reflection and warm absorption features in the source (Nandra \& Pounds 1994; George et al. 2000; Leighly 1999; Ballantyne et al. 2001; Parker et al. 2014; Longinotti et al. 2013, 2019; Ezhikode et al. 2020). Various attempts to model the X-ray spectra of Mrk~335 in the past suggested the possibility of changes in the geometry of the corona leading to the state changes in the source (e.g. Gallo et al., 2013; Wilkins et al., 2015; Gallo et al., 2015;  Gallo L., 2018). Signatures of soft excess has been persistently seen in the X-ray spectra of the source (e.g., Bianchi et al. 2001; Grupe et al. 2001; Grupe et al. 2008; Chainakun \& Young, 2015; Gallo et al. 2015).

The source is also known to show considerable variability in the optical/UV band which was found to be correlated and uncorrelated with the X-ray variability at various epochs (Buisson et al. 2017; Gallo et al. 2018). The long-term monitoring of Mrk~335 with \textit{Swift} in optical--UV--X-ray bands and the observations with other telescopes revealed the properties of the source at different phases of its variability (Buisson et al. 2017; Tripathi et al. 2020).

\begin{table*}
\small
\begin{center}
\caption{The details of \textit{AstroSat} observations of Mrk~335. The quoted count rates are background subtracted values in the  0.3--8~keV band for SXT and in the 4--20~keV band for LAXPC~20 (LXP~20).}
\begin{tabular}{ccccccc}
\toprule
\multicolumn{3}{c}{Observation} & \multicolumn{2}{c}{Exposure (ks)}	& \multicolumn{2}{c}{Count Rate (counts/s)}\\
\cmidrule(l){1-3} \cmidrule(l){4-5} \cmidrule(l){6-7}
Number	& 	ID			&	Date			&	SXT	&	LXP~20			& SXT (10$^{-2}$)	& LXP~20
\vspace*{0.2em}\\
\midrule
\vspace*{0.1em}
Obs~1	&	9000001654	& 31/10/2017		& 14		&	17		&	3.60$\pm$0.33	&	1.14$\pm$0.07 
\vspace*{0.5em}\\
Obs~2	&	9000001700	& 18/11/2017  	& 16		&	24		&	4.14$\pm$0.32	&	1.36$\pm$0.07
\vspace*{0.1em}\\
\bottomrule
\end{tabular}
\end{center}
\label{tab_obs_xray}
\end{table*}

\textit{AstroSat} (Singh et al. 2014, Agrawal 2017) monitored Mrk~335 at two epochs in 2017. We present the results from these multi-wavelength observations in X-ray and UV bands. We studied the X-ray spectral features by modelling the soft and hard X-ray spectra with different models. The details of observations and data processing are given in ${\S}$\ref{obs}. The analysis of X-ray and UV data are described in ${\S}$\ref{xray}, ${\S}$\ref{uvit} and ${\S}$\ref{sed}. In ${\S}$\ref{rslt}, we discuss the results of the study. The summary and discussion of the work are given in {\S}\ref{dscsn_sum}.

\section{Observations}
\label{obs}

Mrk~335 was observed simultaneously in the X-ray and UV bands with Soft X-ray Telescope (SXT: Singh et al., 2017), Large Area X-ray Proportional Counter (LAXPC: Yadav et al. 2016; Antia et al. 2017), Cadmium Zinc Telluride Imager (CZTI: Rao et al. 2017; Bhalerao et al. 2017; Vadawale et al. 2016), and Ultra Violet Imaging Telescope (UVIT: Tandon et al. 2017(a,b)) onboard \textit{AstroSat} on October 31, 2017 (Obs~1) and November 18, 2017 (Obs~2). Here, we use the data from SXT, LAXPC, and UVIT observations. The details of these observations are given in Table~\ref{tab_obs_xray} and Table~\ref{tab_obs_uv}, in the subsequent sections. The data used for the study are available at the Astrobrowse archive handled by Indian Space Science Data Centre (ISSDC). 

\section{X-ray Analysis}
\label{xray}

\subsection{Data reduction}
\label{xray_data}

The Level-2 data products for SXT and LAXPC observations were obtained from the Level-1 data using the processing pipelines. The SXT observations were performed in the photon counting (PC) mode. We used {\sc sxtpipeline~1.4b} (Release Date: 2019-01-04) for reducing the Level-1 SXT data. The pipeline produced cleaned event list for each orbit. These event lists were then merged using the {\sc sxtevtmerger} tool in Julia. The merged event list in each observation was used to create high-level science products using {\sc xselect}. We used the software {\sc LaxpcSoft} for processing LAXPC data. From the Level-2 event file and the GTI file created, we generated the light curves and spectra with the various tasks in the tool. Since the exposure time for the observations are less than that necessary for obtaining a good signal to noise data, and the source was in a low flux state, the data quality is found to be poor.

\begin{figure*}
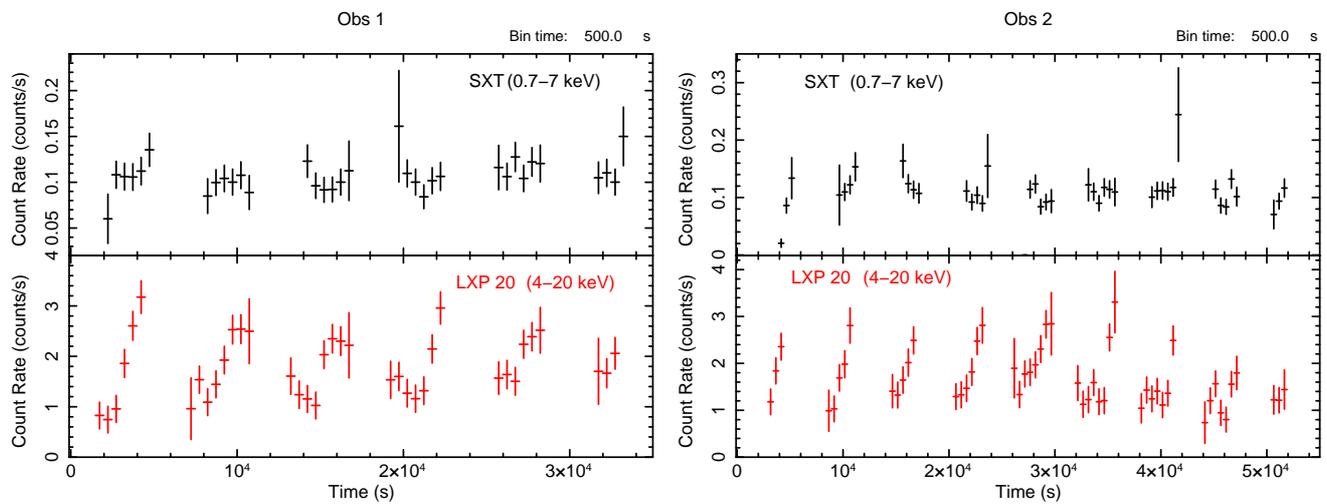

\begin{center}
\includegraphics[trim = 0.2cm 1cm 0.9cm 0.8cm, clip=True, scale=0.36, angle=-90]{Fig1.ps}
\includegraphics[trim = 0.2cm 1cm 0.9cm 0.8cm, clip=True, scale=0.36, angle=-90]{Fig2.ps}
\caption{X-ray light curves for the two observations (Obs~1 \& Obs~2) binned for 500~s. SXT (black) and LAXPC~20 (LXP~20: red) light curves are extracted from 0.7--7~keV and 4--20~keV bands, respectively.}
\label{fig_lc}
\end{center}
\end{figure*}

\subsection{Light curves}
\label{xray_lc}

We created the light curves in both soft and hard X-ray bands. SXT light curves for the two observations were generated in the 0.7$-$7~keV band, from a region of 16~arcmin radius circle, for different time bins using {\sc xselect}. We also generated LAXPC light curves in the energy range of 4--20~keV for various bin sizes. Since the source is very faint in the hard X-ray band, we used the specific {\sc laxpcsoft} code for faint source background for light curve generation. Fig.~\ref{fig_lc} shows the 0.7--7~keV SXT and 4--20~keV LAXPC~20 light curves for the two observations, created for a time bin of 500~s. We checked the variability of the light curves using the ftool {\sc lcstats} and found no variability in SXT. Though the LAXPC light curves showed significant variability in terms of the fraction RMS amplitude, a similar variability pattern was observed in the background lightcurve. Hence, the variability seen in the net LAXPC light curves in Fig.\ref{fig_lc} is not intrinsic to the source.

\begin{figure*}
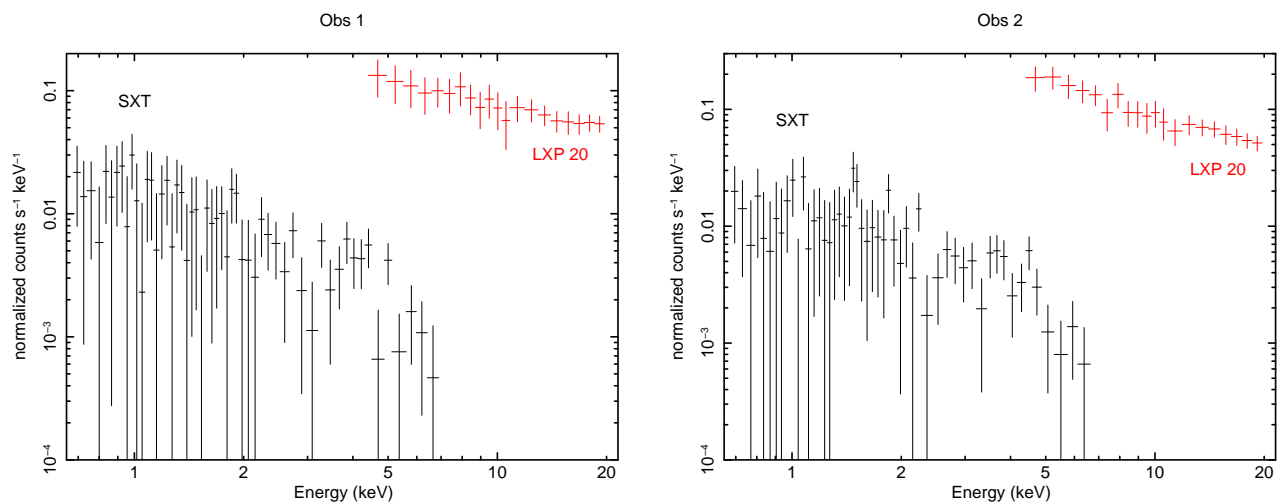

\begin{center}
\includegraphics[scale=0.34,angle=-90]{Fig3.ps}
\includegraphics[scale=0.34,angle=-90]{Fig4.ps}
\caption{\small SXT (0.7--7~keV: black) and LAXPC~20 (4--20~keV: red) spectra of Mrk~335 at the two epochs, Obs~1 and Obs~2.}
\label{fig_spec}
\end{center}
\end{figure*}

\begin{table*}
\tabcolsep 0.5cm
\footnotesize
\begin{center}
\caption{The details of UVIT observations at the two epochs. The last two columns show the measured count rate (background subtracted) from a circular regions of 30 subpixel radius.}
\begin{tabular}{cccccccc}
\toprule
Band		& 	Filter	& Wavelength & Width  & \multicolumn{2}{c}{Exposure (s)}  & \multicolumn{2}{c}{Count Rate (counts/s)} \\
\cmidrule(l){5-6} \cmidrule(l){7-8}
		&			& ($\AA$) & ($\AA$)& Obs~1 & Obs~2 & 	Obs~1 & Obs~2 \\
\midrule
\multirow{4}{*}{NUV}
& N242W & 2418.0 & 785.0 & 1881.957 & 1949.968 & 27.30 $\pm$ 0.12 & 27.81 $\pm$ 0.12 \\
& N245M & 2447.0 & 280.0 & 667.5162 & 2410.008 & 15.35 $\pm$ 0.15 & 16.16 $\pm$ 0.08 \\
& N263M & 2632.0 & 275.0 & 2523.842 & 1361.323 & 13.35 $\pm$ 0.07 & 13.97 $\pm$ 0.10 \\
& N279N & 2792.0 & 90.0 & 563.2253 & 688.3519 & 3.48 $\pm$ 0.08 & 3.37 $\pm$ 0.07 \\
\midrule
\multirow{4}{*}{FUV} 
& F148Wa & 1485.0 & 500.0 & 1874.5 & 816.4949 & 10.70 $\pm$ 0.08 & 11.67 $\pm$ 0.12 \\
& F154W & 1541.0 & 380.0 & 1025.182 & 761.7113 & 8.64 $\pm$ 0.09 & 9.54 $\pm$ 0.11 \\
& F169M & 1608.0 & 290.0 & 1876.111 &  & 6.76 $\pm$ 0.06 &  \\
& F172M & 1717.0 & 125.0 & 564.0201 &  & 2.24 $\pm$ 0.06 &  \\
\bottomrule
\end{tabular}
\label{tab_obs_uv}
\end{center}
\end{table*}

\subsection{Spectral Analysis}
\label{xray_spec}

The SXT source spectra were extracted from circular regions of 16~arcmin radius, whereas the blank sky spectrum was used for the background. We used the background spectrum and the response files provided by the SXT-POC team. The Ancillary Response Function (ARF) file corrected for vignetting, PSF and exposure was generated using the latest module released on 2019 July 18, {\sc sxteefmodule{\_}v02}. The rmf file for grade 0--12 was used for the analysis. The spectra were also grouped so that we can apply $\chi^2$ statistic. As the SXT response is not well characterised below 0.7~keV, the region was ignored in the analysis.

The LAXPC background spectra were created with the faint source code mentioned above. Since the background is more stable for LAXPC~20, we used only LAXPC~20 spectra for the analysis. Here, we ignored below 4~keV and above 20~keV as the regions were dominated by background. The SXT and LAXPC spectra at the two epochs are shown in Fig.~\ref{fig_spec}.

Both the SXT and LAXPC~20 spectra were analysed simultaneously to characterise the broadband continuum of the source. The spectral analysis was done using {\sc xspec} version~12.9. To account for the shift in SXT response, \emph{gain} command in {\sc xspec} was used with \textit{offset} parameter fixed at 0.02. The model {\sc constant} was used to take care of the cross normalisation between SXT and LAXPC~20. Also, a systematic error of 3\% is applied while fitting.

The X-ray spectral analysis was started by jointly fitting the SXT and LAXPC~20 spectra in the hard X-ray band (2$-$20~keV) with an absorbed power-law (\textit{tbabs$\times$powerlaw}) model. The Galactic column density ($N_{\rm H}$) for the \textit{tbabs} component was fixed at 3.56$\times$10$^{20}$~cm$^{-2}$ obtained from the LAB survey (Kalberla et al. 2005). The fit yielded a photon index ($\Gamma$) of less than 1 for both the observations. To check the presence of intrinsic absorption, we added a \textit{ztbabs} component. However, the fit did not improve, and the intrinsic equivalent Hydrogen column density ($N_{\rm H}^{\rm Int}$) was not constrained. Since such flat hard X-ray spectra could be the result of intrinsic absorption and the presence of distant reflection, we also included one \textit{xillver} (Garcia et al. 2010, 2013) component. Only the normalisation ($N_{\rm xl}$) and reflection fraction ($f_{\rm refl}$) parameters of \textit{xillver} were allowed to vary during the fit. The photon index of \textit{xillver} component was tied to the slope of \textit{powerlaw} model. Inclination (\textit{i}), high-energy cut-off ($E_{\rm cut}$) and the iron abundance ($A_{\rm Fe}$) were fixed at 30$^\circ$, 300~keV and 1 (in solar abundance), respectively. The ionisation parameter ($\xi$) was set to the minimum value, log$\xi$=0, to account for the reflection from neutral material. The new model yielded a marginally better fit for both the observations with $\Delta \chi^2 \sim -4.9$ for Obs~1 and $\Delta \chi^2 \sim -8.5$ for Obs~2 for a change in degrees of freedom (dof) of 2. The photon index also slightly increased with the addition of \textit{xillver} component.

Further, we noticed the energy range below 2~keV and found that the spectrum rises above the current model. This is a clear indication of the soft excess emission. Therefore, we added \textit{bbody} to model the soft X-ray excess and the fit provided a blackbody temperature ($kT_{\rm bb}$) of $\sim 0.1$~keV. All parameters are well constrained, except for the \textit{xillver} normalisation and reflection fraction. However, the model $tbabs(ztbabs\times(powerlaw+bbody)+xillver)$ was preferable than the one without either \textit{xillver} or \textit{bbody} component.

\begin{figure*}
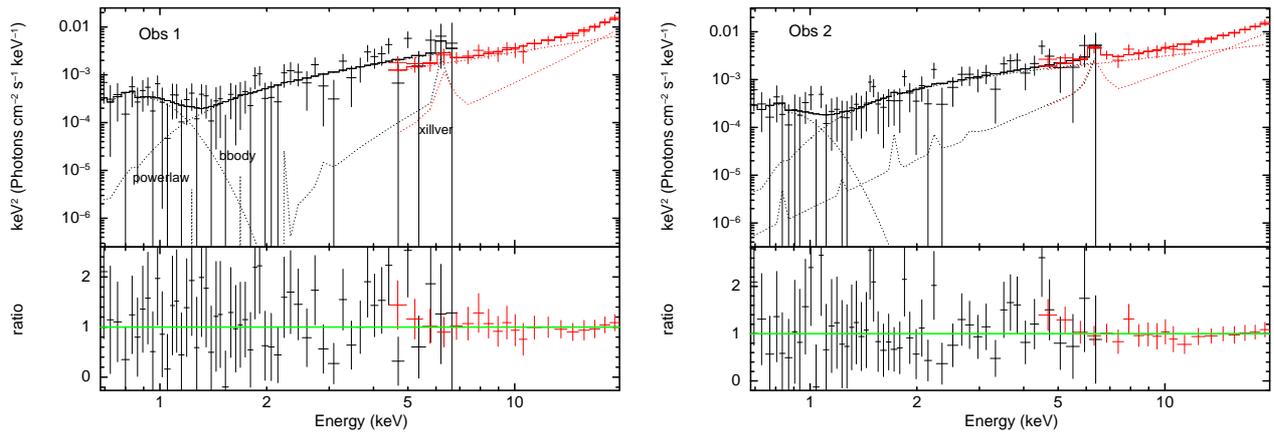

\begin{center}
\includegraphics[scale=0.32,angle=-90]{Fig5.ps}
\includegraphics[scale=0.32,angle=-90]{Fig6.ps}
\caption{The Spectral fitting plots for the model $tbabs(ztbabs{\times}(powerlaw+bbody)+xillver)$ in the 0.7--20~keV for the two observations. The unfolded spectra and the model are shown in the upper panels and the ratio of data to model are shown in the lower panels. The solid lines in the upper panels represent the overall model whereas the dotted lines deonte the individual model components \textit{powerlaw}, \textit{bbody} and \textit{xillver}. The SXT data and model are shown in black colour while those of LAXPC are shown in red colour.}
\label{fig_fit}
\end{center}
\end{figure*}

Another possible reason behind the observed hard spectrum could be the presence of partial covering absorption. Therefore, we also tried fitting the spectra with \textit{zpcfabs} and \textit{zxipcf} models. However, the fits provided poor statistic and poorly constrained parameters.

\section{UVIT Analysis}
\label{uvit}

The source was observed with UVIT at both NUV and FUV wavelengths. Four filters (in PC mode) were used for both NUV and FUV observations at the first epoch. In the second observation, only two FUV filters were used as the instrument stopped working during that time. The filter information and other details of the exposures are given in Tabel~\ref{tab_obs_uv}. The Level-2 data, already processed with the latest pipeline {\sc UVIT Level-2 Pipeline (UL2P)} version~6.3 by the POC were available at the archive. We used these data sets for further analysis.

We carried out the photometry on the combined image in each filter. The combined images are obtained from the Level-2 data created using aspect correction done with VIS or NUV data. To do the photometry, we chose a circular region of radius 30~subpixels ($\sim$12.5") centred around the source in each filter. This would give enclosed energy of around 97\% in both NUV and FUV filters (Tandon et al. 2020). The observed count rates are given in Table~\ref{tab_obs_uv}. Most of these observations suffer from saturation. Hence, we followed the procedure described in Tandon et al. (2017, 2020) to correct for the effect. We note that the NUV N242W filter records a total count rate of $\sim$30~counts/s where the above-mentioned saturation correction is not valid. Hence, we do not use the data from N242W filter for further analysis. The background regions were selected from different circles with radii of 60~subpixels. The average background count rates were then subtracted from the saturation corrected values. We obtained the AB magnitude ($m_{\rm AB}$) from the count rates and the zero-point (ZP) magnitudes and calculated the corresponding flux density $F_\lambda$ (Tandon et al., 2017, 2020) in each filter. We have also applied the Galactic extinction correction for the estimated $F_\lambda$ values using Cardelli et al. (1989) relation for R$_V$=3.1 and A$_V$=0.118 (Schlegel et al., 1998).

\section{Spectral Energy Distribution}
\label{sed}

A comprehensive modelling of the UV to X-ray spectral energy distribution (SED) of AGN can unveil the geometry of the central emitting regions and physics related to the variability mechanisms. We have UV observations of Mrk~335 with various filters in NUV and FUV bands (see Table~\ref{tab_obs_uv}). But modelling the accretion disc emission from these photometric data is complicated as many components, like host galaxy and emission lines, can contribute to the observed flux at these wavelengths. Moreover, most of these observations are affected by saturation. Though we corrected the source count rates for the saturation effects (as mentioned in the previous section), there could still be uncertainties associated with the flux estimation. Also, it is difficult to derive the source flux free from the host galaxy contamination. To account for these uncertainties, we added a systematic of 5--10\% and fitted the UV--X-ray SED. The UV spectra were created by converting the flux (corrected for the Galactic reddening) in each filter (given in Table.~\ref{tab_flux_uv}) using the task {\sc flx2xsp}.

\begin{figure}[!tb]
\begin{center}
\includegraphics[scale=0.33,angle=-90]{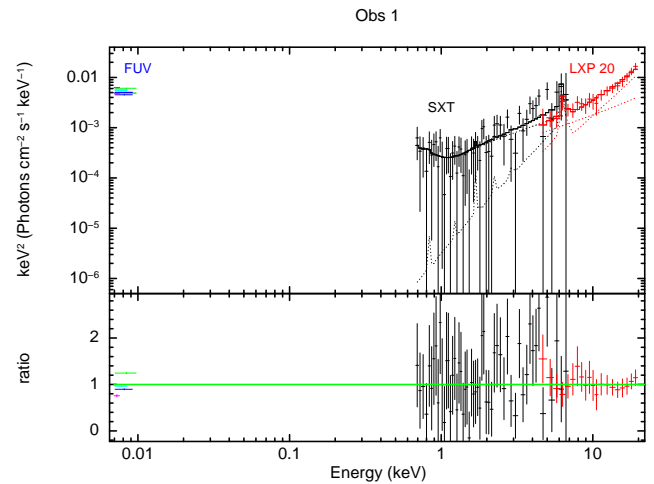}
\caption{Broadband FUV--X-ray SED of Mrk~335 with SXT, LAXPC~20 and UVIT (FUV filters: F148Wa, F154W, F169M \& F172M) data from Obs~1. The data are modelled with \textit{optxagnf} and \textit{xillver}, modified by both Galactic and intrinsic absorption and reddening.}
\label{fig_sed}
\end{center}
\end{figure}

We used the {\sc xspec} model \textit{optxagnf} (Done et al., 2012) to fit the broadband SED of the source. The model can describe the emissions form accretion disc together with the soft and hard X-ray components. Here, we show the example of SED fitting for Obs~1 since there are more FUV data points for this observation. We began with the analysis of X-ray spectra by replacing \textit{powerlaw+bbody} components with \textit{optxagnf} in the best-fit model $tbabs(ztbabs{\times}(powerlaw+bbody)+xillver)$. Further, we added the FUV  spectra for the filters F148Wa, F154W, F169M \& F172M, and included the model \textit{zreddedn} to correct for intrinsic reddening. The parameter E(B-V) for \textit{zredden} was obtained from the intrinsic column density $N_{\rm H}^{\rm Int}$ using the relation given by Bessell (1991). The parameters of \textit{xillver} for the X-ray part were fixed at the best-fit values, whereas the component was not used for UV spectra.  The cross normalisation constant for both the NUV and FUV spectral groups were tied to that of the SXT spectrum. We notice that fitting the FUV--X-ray SED resulted in a reasonable $\chi^2$ of 91.78 for 71 degrees of freedom when a systematic error of 5\% was applied. The fit provided a hard X-ray photon index of $\sim$1.1 and intrinsic $N_{\rm H}$ of ${\rm 2.5 \times 10^{21} cm^{-2}}$ (fixed). The other parameters obtained from the fit are Eddington ratio $\sim$1, coronal radius $R_{\rm cor} \sim 19~R_g$ and the fraction of power below $R_{\rm cor}$ emitted as the hard X-ray component $f_{pl} \sim 0.9$. The temperature and optical depth of the soft X-ray component are $\sim$0.1~keV, $\sim$79.4, respectively. The SED plot for the fit is shown in Fig.~\ref{fig_sed}. When the NUV spectra were added, the fit worsened. The fit seemed to be improving when the systematic error was increased up to 10\%. From FUV--X-ray SED analysis, we see that the accretion disc is truncated at a radius of about $19~R_g$, below which the energy is dissipated as Comptonised emission. However, this is a preliminary analysis and the errors on parameters are not obtained. A detailed and systematic study of the multi-wavelength SED of Mrk~335 with \textit{AstroSat} data will be done later.

\section{Results}
\label{rslt}

\begin{table*}
\tabcolsep 0.5cm
\footnotesize
\begin{center}
\caption{Best-fit parameters for the spectral fits in the 0.7--20~keV band for \textit{AstroSat} (SXT \& LAXPC~20) observations taken on 31/10/2017 (Obs1) and 18/11/2017 (Obs2). The \textit{xillver} parameters that kept fixed while fitting are $i=30^\circ$, $A_{\rm Fe}$=1, log$\xi=0$, and $E_{\rm cut}$=300~keV.}
\begin{tabular}{c|cccc}
\toprule
Energy	& \multirow{1}{*}{Model} 	& \multirow{1}{*}{Parameter}	& \multicolumn{1}{c}{Obs~1}	& \multicolumn{1}{c}{Obs~2} \\
Range	&	& & 		 \vspace{0.4em}\\
\midrule
\multirow{20}{*}{2--20~keV}	&	$tbabs{\times}powerlaw$	& $\Gamma$					& 0.31$^{+0.28}_{-0.32}$		& 0.66$^{+0.22}_{-0.23}$
\vspace{0.4em}\\
&						& $N_{\rm pl}$ (10$^{-4}$)	& 1.64$^{+0.79}_{-0.60}$		& 2.69$^{+0.96}_{-0.77}$
\vspace{0.4em}\\
&						& constant					& 0.50$^{+0.31}_{-0.20}$		& 0.87$^{+0.40}_{-0.28}$
\vspace{0.4em}\\
&\multicolumn{1}{c}{$\chi^2$/dof}				&		& 28.74/39					& 31.68/38 \vspace{0.8em}\\
& $tbabs{\times}ztbabs{\times}powerlaw$ & $N_{\rm H}^{\rm Int}$ (10$^{22}$cm$^{-2}$) & $<3.7$	& $<4.1$
\vspace{0.4em}\\
&						& $\Gamma$					& 0.35$^{+0.31}_{-0.33}$		& 0.74$^{+0.24}_{-0.26}$
\vspace{0.4em}\\
&						& $N_{\rm pl}$ (10$^{-4}$)	& 1.88$^{+1.74}_{-0.78}$		& 3.73$^{+2.70}_{-1.58}$
\vspace{0.4em}\\
&						& constant					& 0.49$^{+0.30}_{-0.19}$		& 0.79$^{+0.37}_{-0.24}$
\vspace{0.4em}\\
&\multicolumn{1}{c}{$\chi^2$/dof}		&				& 28.56/38					& 30.28/37 \vspace{0.8em}\\
& $tbabs(ztbabs{\times}powerlaw+xillver)$ & $N_{\rm H}^{\rm Int}$ (10$^{22}$cm$^{-2}$) & $<4.8$	& 2.43$^{+2.94}_{-2.12}$
\vspace{0.4em}\\
&						& $\Gamma$					& 0.85$^{+0.41}_{-0.26}$		& 1.27$^{+0.39}_{-0.37}$
\vspace{0.4em}\\
&						& $N_{\rm pl}$ (10$^{-4}$)	& 3.91$^{+2.59}_{-1.39}$		& 6.89$^{+4.61}_{-2.96}$
\vspace{0.4em}\\
&						& $f_{\rm refl}$ 			& $>0.007$					& $>$0.7	
\vspace{0.4em}\\
&						& $N_{\rm xl}$ (10$^{-5}$)	& $>158.7$					& $>$1.7
\vspace{0.4em}\\
&						& constant					& 0.55$^{+0.31}_{-0.19}$		& 0.82$^{+0.37}_{-0.24}$
\vspace{0.4em}\\
&\multicolumn{1}{c}{$\chi^2$/dof}		&				& 23.66/36					& 21.83/35 \vspace{0.4em}
\\
\midrule
\\
\multirow{28}{*}{0.7--20~keV}	&	$tbabs(ztbabs{\times}powerlaw+xillver)$ & $N_{\rm H}^{\rm Int}$ (10$^{22}$cm$^{-2}$) & 	$<0.3$	& $<0.9$
\vspace{0.4em}\\
&										 & $\Gamma$					& 0.83$^{+0.41}_{-0.28}$ & 1.05$^{+0.45}_{-0.32}$
\vspace{0.4em}\\
&						& $N_{\rm pl}$ (10$^{-4}$)	& 2.56$^{+0.71}_{-0.49}$		& 3.74$^{+2.07}_{-1.20}$
\vspace{0.4em}\\
&						& $f_{\rm refl}$ 			& $>0.003$					& 0 -- 0(?)
\vspace{0.4em}\\
&						& $N_{\rm xl}$ (10$^{-5}$)	& $>0.3$						& $>2.5$	
\vspace{0.4em}\\
&						& constant					& 0.79$^{+0.24}_{-0.30}$		& 1.03$^{+0.48}_{-0.30}$
\vspace{0.4em}\\
&\multicolumn{1}{c}{$\chi^2$/dof}		&				& 47.57/68					& 45.28/66 \vspace{0.8em}\\
& $tbabs(ztbabs{\times}(powerlaw+bbody)+xillver)$	& $N_{\rm H}^{\rm Int}$ (10$^{22}$cm$^{-2}$) & 0.89$^{+0.98}_{-0.70}$		& 0.86$^{+0.78}_{-0.64}$
\vspace{0.4em}\\
&						& $\Gamma$					& 0.85$^{+0.39}_{-0.26}$		& 1.18$^{+0.40}_{-0.36}$
\vspace{0.4em}\\
&						& N$_{pl}$ (10$^{-4}$)		& 3.47$^{+1.50}_{-0.99}$	 	& 5.04$^{+2.68}_{-1.87}$
\vspace{0.4em}\\
&						& $kT_{bb}$ 	(keV)			& 0.08$^{+0.04}_{-0.02}$		& 0.07$^{+0.03}_{-0.06}$
\vspace{0.4em}\\
&						& N$_{bb}$ (10$^{-3}$)		& 4.23$^{+39.04}_{-4.17}$	& 6.66$^{+143.85}_{-6.60}$
\vspace{0.4em}\\
&						& $f_{\rm refl}$ 			& 0.06$^{+999.92}_{-0.02}$	& 5.16$^{+994.84}_{-4.29}$
\vspace{0.4em}\\
&						& N$_{xl}$	 (10$^{-4}$)		& $>8.0$						& $>1.6$
\vspace{0.4em}\\
&						& constant					& 0.60$^{+0.29}_{-0.19}$		& 0.95$^{+0.41}_{-0.28}$
\vspace{0.4em}\\
&\multicolumn{1}{c}{$\chi^2$/dof}		&			& 39.46/66					& 39.37/64 \vspace{0.8em}\\
& $tbabs{\times}ztbabs{\times}(powerlaw+bbody)$	& $N_{\rm H}^{\rm Int}$ (10$^{22}$cm$^{-2}$) & $<1.2$ & $<1.1$
\vspace{0.4em}\\
&						& $\Gamma$					&  0.34$^{+0.29}_{-0.32}$	& 0.68$^{+0.23}_{-0.12}$
\vspace{0.4em}\\
&						& N$_{pl}$ (10$^{-4}$)		& 1.81$^{+1.10}_{-0.73}$		& 2.92$^{+1.43}_{-0.05}$
\vspace{0.4em}\\
&						& $kT_{bb}$ 					& 0.10$^{+0.08}_{-0.04}$		& 0.08$^{+0.08}_{-0.07}$
\vspace{0.4em}\\
&						& N$_{bb}$ (10$^{-3}$)		& 1.40$^{+42.11}_{-1.36}$	& 0.43$^{+14.03}_{-0.43}$
\vspace{0.4em}\\
&						& constant							& 0.49$^{+0.29}_{-0.20}$		& 0.85$^{+0.39}_{-0.27}$
\vspace{0.4em}\\
&\multicolumn{1}{c}{$\chi^2$/dof}			&			& 43.76/68					& 46.21/66	\vspace{0.4em}\\
\bottomrule
\end{tabular}
\label{xray_par}
\end{center}
\end{table*}

\begin{table*}
\tabcolsep 0.5cm
\footnotesize
\begin{center}
\caption{The total unabsorbed X-ray flux for the model $tbabs(ztbabs{\times}(powerlaw+bbody)+xillver)$ from the two \textit{AstroSat} observations and the near-XRT observation.}
\begin{tabular}{cccc}
\toprule
Energy range	&	\multicolumn{3}{c}{Flux (${\rm 10^{-11}~erg~cm^{-2}~s^{-1}}$)} \\
\cmidrule(l){2-4}
(keV)		& Obs~1						&	Obs~2					& XRT \\
\midrule
0.7 -- 20	& 2.14 $^{+0.86}_{-0.63}$	& 1.66 $^{+0.59 }_{-0.45}$	& \\
0.7 -- 2		& 0.06 {\tiny $\pm0.01$}		& 0.06 {\tiny $\pm0.01$}		& \\
2 -- 20		& 2.09 $^{+0.86}_{-0.62}$	& 1.61 $^{+0.59 }_{-0.45}$	& \\
2 -- 10		& 0.64 $^{+0.18}_{-0.16}$	& 0.59 $^{+0.15 }_{-0.14}$	& 0.32 $^{+0.21}_{-0.14}$ \\
0.3 -- 2		& 0.07 $^{+0.07 }_{-0.02}$	& 0.07 $^{+0.25 }_{-0.02}$	& 0.16 $^{+0.06}_{-0.05}$ \\
0.3 -- 10	& 0.71 $^{+0.17 }_{-0.15}$	& 0.66 $^{+0.09 }_{-0.13}$	& 0.49 $^{+0.21}_{-0.15}$ \\
\bottomrule
\end{tabular}
\label{xray_flux}
\end{center}
\end{table*}

As expalined in Sec.~\ref{xray_spec}, we tried fitting the X-ray spectra with different models. The best-fit parameters for these models are given Table~\ref{xray_par}. Fig.~\ref{fig_fit} shows the various spectral fitting plots for Obs~1 and Obs~2. We found that \textit{tbabs(ztbabs$\times$(powerlaw+bbody) + xillver)} better fits the data in the 0.7--20~keV band than the other models. We estimated the unabsorbed flux in different energy bands using the \textit{cflux} convolution model. The obtained flux values are provided in Table~\ref{xray_flux}.

Both observations are found to be intrinsically absorbed with $N_{\rm H}^{\rm Int} \sim 9\times 10^{21} {\rm cm}^{-2}$. The source does not show significant X-ray spectral or flux variability between the observations. The X-ray spectra seem to be harder in both the observations with $\Gamma \sim 0.9 (1.2)$ (consistent within errorbars) for Obs~1(Obs~2) while the reflection parameters are not properly constrained. The total flux in the 2--20~keV is roughly ${\rm 2 \times 10^{-11}~erg~cm^{-2}~s^{-1}}$, and in the 0.7--2~keV band it is around ${\rm 6 \times 10^{-13}~erg~cm^{-2}~s^{-1}}$.

\begin{table*}
\footnotesize
\begin{center}
\caption{Results from the analysis of UVIT data. The saturation corrected net count rates with the corresponding magnitude and flux density ($F_\lambda$) for each filter are given in the last tree columns. $F_\lambda$ values are corrected for the Galactic reddening.}
\begin{tabular}{cccccccc}
\toprule
Filter	&	 Zero Point			& \multicolumn{2}{c}{Count Rate (counts/s)} & \multicolumn{2}{c}{$m_{\rm AB}$ (magnitude)} & \multicolumn{2}{c}{$F_\lambda (10^{-14} {\rm erg~cm^{-2}s^{-1}})$}  \\
\cmidrule(l){3-4} \cmidrule(l){5-6} \cmidrule(l){7-8}
		&	(magnitude)		& Obs~1			   & Obs~2		& Obs~1			& Obs~2			& Obs~1		& Obs~2	\\		
\midrule
N245M & 18.452+/-0.005 & 21.84 $\pm$ 0.18 & 23.51 $\pm$ 0.1 & 15.104+/-0.010 & 15.024+/-0.007 & 2.15 $\pm$ 0.02 & 2.32 $\pm$ 0.01 \\
N263M & 18.146+/-0.010 & 18.14 $\pm$ 0.08 & 19.3 $\pm$ 0.12 & 14.999+/-0.011 & 14.932+/-0.012 & 1.98 $\pm$ 0.02 & 2.1 $\pm$ 0.02 \\
N279N & 16.416+/-0.010 & 3.75 $\pm$ 0.08 & 3.62 $\pm$ 0.07 & 14.982+/-0.026 & 15.020+/-0.024 & 1.76 $\pm$ 0.04 & 1.7 $\pm$ 0.04 \\
\midrule
F148Wa& 17.994 $\pm$ 0.010 & 13.24 $\pm$ 0.08 & 14.78 $\pm$ 0.13 & 15.189+/-0.012 & 15.070+/-0.014 & 5.55 $\pm$ 0.06 & 6.2 $\pm$ 0.08 \\
F154W & 17.771 $\pm$ 0.010 & 10.2 $\pm$ 0.1 & 11.5 $\pm$ 0.12 & 15.250+/-0.015 & 15.119+/-0.015 & 4.84 $\pm$ 0.07 & 5.46 $\pm$ 0.08 \\
F169M & 17.410 $\pm$ 0.010 & 7.67 $\pm$ 0.06 & 			   & 15.198 $\pm$ 0.013 & & 4.63 $\pm$ 0.06 & \\
F172M & 16.274 $\pm$ 0.020 & 2.34 $\pm$ 0.06 & 			   & 15.350 $\pm$ 0.040 & & 3.51 $\pm$ 0.12 & \\
\bottomrule
\end{tabular}
\label{tab_flux_uv}
\end{center}
\end{table*}

\begin{figure}
\begin{center}
\includegraphics[trim = 0cm 0cm 0cm 0cm, clip=True, scale=0.58]{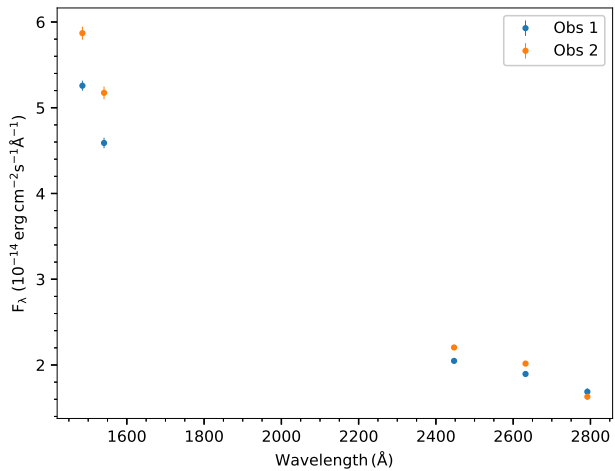}
\caption{\small Plot showing the variability in FUV (F148Wa \& F154W) and NUV (N245M, N263M \& N279M) emissions from Mrk~335 between the two epochs.}
\label{fig:flux_UV}
\end{center}
\end{figure}

Unlike the X-ray observations, UV emission from the source shows variability in both FUV and NUV bands. The net count rate and flux in each filter are mentioned in Table~\ref{tab_flux_uv}. The flux variability (except for the FUV filters F169M, F172M and the NUV filter N245M) are shown in Fig.~\ref{fig:flux_UV}. In order to check if the variability is an instrument artefact, we obtained the light curves of a star in both NUV and FUV images (since the stars were too faint in NUV N279, F154W and F148Wa filters we did not obtain the count rate for those exposures). The net count rates for the star seem to be non-variable in the NUV band showing that the variability shown by the source is real.

\subsection{Comparison with other observations}
\label{swift}

Mrk~335 has been observed at optical, UV and X-ray wavelengths with various missions. Here, we give a brief summary of the analysis of some of these data and compare those with the results from our \textit{AstroSat} observations. \textit{Swift} has been monitoring Mrk~335 for years in X-rays and optical/UV. We analysed one \textit{Swift} observation close to \textit{AstroSat} observations as there are no observations strictly simultaneous with that of \textit{AstroSat}. We retrieved XRT and UVOT data taken on November 03, 2017 (almost three days after Obs~1 and two weeks before Obs~2). This near-simultaneous \textit{Swift} observation (ID: 00033420140) has an exposure time of only $\sim 800$s. The XRT observation was made in PC mode. We reduced the data with {\sc xrtpipeline} and extracted the spectrum and light curve from a circular region of radius 30~pixels. The background region of 50~pixels radius circle was also selected from the same image. We generated the XRT light curve in the 0.3--10~keV band and did not find any variability. The net count rate of the XRT observation is around 0.07 counts s$^{-1}$ (0.3--10~keV). Since the data quality is not good, we grouped the spectrum for minimum 5 counts per bin and used \textit{cstat} while fitting. Modelling the 0.3-10~keV spectrum with $TBabs \times zTBabs \times powerlaw$ gave a photon index of about 1.8, softer than that obtained for the fit of combined SXT and LAXPC spectra in the 0.7--20~keV band. The corresponding fit-static is \textit{cstat}/dof = 14.40/9. When a \textit{xillver} component was added the fit-static reduced to 1.46/7 with a steeper unconstrained photon index. Other parameters are $N_{\rm H}{^{\rm Int}} = 11.75_{-6.37}^{15.94} \times 10^{22} cm^{-2}$, $N_{\rm pl} = 7.59_{-5.61}^{+9.68} \times 10^{-3}$, $N_{\rm xl} = 6.76_{-2.45}^{+2.78} \times 10^{-6}$ and $f_{\rm refl}$ was not constrained. Adding a \textit{bbody}, with $kT_{\rm bb}$ fixed at 0.1~keV did not change the fit-statistic any more. However, the spectrum remained steeper with $\Gamma > 2.3$. We also fitted the spectrum with $TBabs \times zTBabs(bbody+powerlaw)$ model which resulted in a \textit{cstat}/dof of 1.35/8 with parameter values, $N_{\rm H}{^{\rm Int}} < 0.4\times 10^{22} cm^{-2}$, $\Gamma < 1.2$, $N_{\rm pl} = 1.04_{-0.61}^{+1.71}\times 10^{-4}$, $kT_{\rm bb} = 0.14_{-0.06}^{+0.04}$keV, $N_{\rm bb} = 0.02_{-0.01}^{+0.11} \times 10^{-3}$. Here, the blackbody normalisation is lower than the results from \textit{AstroSat} whereas the temperature remains similar. In the optical/UV band, the observation was made with only uvw2 filter. We estimated the flux of the source using the task {\sc uvotproducts}. For this, we selected a source region of 5" radius circle and background circles of larger radius at different regions from the observed images. The background subtracted flux for \textit{uvw2} filter is ${\rm 2.32\pm0.05 \times 10^{-14} erg~cm^{-2}s^{-1}\AA^{-1}}$ (not corrected for Galactic reddening).

The X-ray spectrum of Mrk~335 is complex to be modelled with low-quality data from \textit{AstroSat} and \textit{Swift}. We notice that when the X-ray spectra from XRT and \textit{AstroSat} (SXT \& LAXPC~20) observations were fitted in the same energy range of 0.7--10~keV with simple models like ($tbabs \times zTBabs \times powerlaw$), the parameters agree well within errorbars. The discrepancy arises when we fit the broadband X-ray spectrum, including LAXPC data.

We also carried out a preliminary analysis of one of the \textit{XMM-Newton} observations of Mrk~335 taken in 2019 January. The EPIC-pn spectrum (net exposure $\sim$ 66~ks) showed soft excess emission and a broad iron emission line. We fitted the spectrum in the 0.3--10~keV band with an absorbed (Galactic and intrinsic) power-law, blackbody and a redshifted broad Gaussian component. The fit yielded a photon index of $\sim$0.7 and $N_{\rm H}{^{\rm Int}}<$0.19. When a \textit{xillver} component was added, the fit improved significantly, and the photon index increased to around 1.4 while $N_H$ remained unconstrained. The X-ray (2--10~keV) flux from the observation is found to be decreased roughly by a factor of 5--6 as compared to the \textit{AstroSat} observations, but $\Gamma$ is consistent within error bars.

\section{Discussion \& Summary}
\label{dscsn_sum}

We found harder X-ray spectra for both \textit{AstroSat} observations of Mrk~335. A similar photon index was obtained when the XRT spectrum was fitted with an absorbed power-law and blackbody model (0.3--10~keV), that is consistent with the result obtained by Tripathi et al. (2020). However, when a reflection component was added in the model, the primary power-law appeared to be softer in the XRT observation. The significance of reflection in the source was noticed in the previous studies as well. For example, Parker et al. (2019) studied the \textit{XMM-Newton}, \textit{Swift} and \textit{NuSTAR} spectra taken in 2018--2019 when the source was showing an extremely low flux level in X-rays. By modelling the broadband continuum in detail, they found that the hard X-ray spectrum is dominated by distant reflection and the soft part by photoionised emission lines. They also observed steep X-ray spectra $\Gamma \sim 2$ and a significant blackbody component. Earlier observations of Mrk~335 reported partial covering absorption and relativistic reflection in the source (Longinotti et al. 2019, Parker et al. 2019) that we could not parameterise with the \textit{AstroSat} data. In a previous study on the correlation between the reflection fraction and photon index in AGN (Ezhikode et al., 2020), we analysed the \textit{NuSTAR} spectrum of Mrk~335 observed in 2013 June. The spectrum showed the presence of broad and narrow emission lines, and we fitted the 3--79~keV spectrum using the models \textit{relxill} and \textit{xillver}. The spectrum was steep with gamma around 2.2, and we did not see any significant intrinsic absorption.

\begin{figure}
\begin{center}
\includegraphics[scale=0.32,angle=-90]{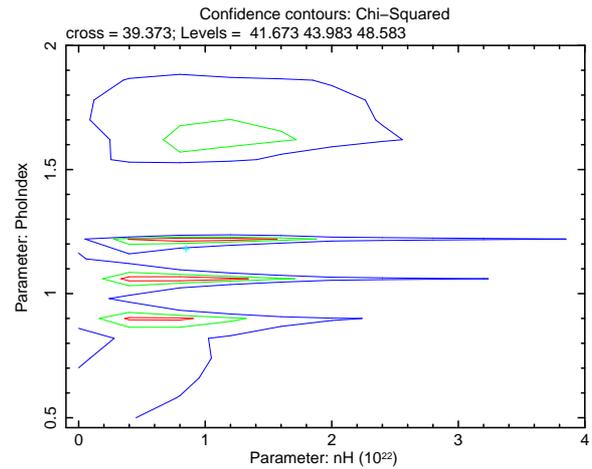}
\caption{The confidence (one, two and three sigma) contour plot of the parameters $N_{\rm H}^{\rm Int}$ and $\Gamma$ for Obs~2.}
\label{fig_contour}
\end{center}
\end{figure}

A larger X-ray photon index of $\Gamma\gtrsim1.5$ is typically observed in NLS1s. Here, we observe a different behaviour even after including the neutral reflection model (though the slope of the second observation is marginally within this range). We also note that the X-ray spectrum gets steeper when the intrinsic obscuration is fixed at larger values, although the fit worsens. We obtained the confidence contour plot (see Fig.~\ref{fig_contour}) for photon index and intrinsic absorption. It is clear from the plot that the data could not constrain $\Gamma$ well and a higher index similar to those found in other AGN is not ruled out. The X-ray emission from the source may be obscured intrinsically, and hence the distant reflection could be dominating the observed spectra.

During \textit{AstroSat} observations, Mrk~335 was in a low-flux state in the UV band as well. Our observations, separated by almost 18~days, show variability in both NUV fand FUV emissions. However, no significant variability was found between the two X-ray observations. Variable UV emission  using \textit{Swift} UVOT observations was detected by Grupe et al. (2008) on time-scales of days to weeks. They found the UV variability to be following the XRT light curve, suggesting the possibility of the same mechanism triggering both the emissions. 

Considering the time-scale of variability in our observations, X-ray reprocessing could be the origin of the observed UV variability in the source. However, a similar variability is not observed in X-ray emissions, and we do not have enough monitoring observations to confirm this. The obscuration of X-rays by clouds could be another possibility of the observed UV variability that is unrelated to X-ray emission. Detailed modelling of the broadband SED is required to explain the scenario. Owing to the low signal-to-noise of the X-ray spectra and uncertainties in the UV flux measurements, a proper modelling of the UV-X-ray SED is difficult. Hence, we do not make a definitive statement regarding the UV variability in the source. A more detailed characterisation of broadband X-ray continuum emissions may be carried out with future better observations with \textit{AstroSat}. With simultaneous filter and grating observations with UVIT, we can also study the nature of variability in the accretion disc emission in depth.


%

\section*{Acknowledgements}
\label{ack}


We would like to acknowledge the anonymous referee for the helpful comments and suggestions. We thank Prof. Shyam Tandon and Mr. Prajwel Joseph for the useful discussions on UVIT data analysis. This publication uses the data from the \textit{AstroSat} mission of the Indian Space Research Organisation (ISRO), archived at the Indian Space Science Data Centre (ISSDC). This work has used the data from the Soft X-ray Telescope (SXT) developed at TIFR, Mumbai, and the SXT POC at TIFR is thanked for verifying and releasing the data via the ISSDC data archive and providing the necessary software tools. We thank the UVIT POC at IIA, Bangalore for the data and their support. This research has made use of data, software and/or web tools obtained from the High Energy Astrophysics Science Archive Research Center (HEASARC), a service of the Astrophysics Science Division at NASA/GSFC and of the Smithsonian Astrophysical Observatory's High Energy Astrophysics Division.



\balance

\end{document}